\documentclass[twocolumn,showpacs,preprintnumbers,amsmath,amssymb,prl]{revtex4-1}
\usepackage{graphicx}
\usepackage{multirow}
\usepackage{bm}
\usepackage[breaklinks=true,colorlinks=true,linkcolor=blue,urlcolor=blue,citecolor=blue]{hyperref}
\usepackage{booktabs}

\graphicspath{{./figures/}}

\begin{document}
\title{Giant tunneling magnetoresistance based on spin-valley-mismatched ferromagnetic metals}

\author{Kun \surname{Yan}$^{1}$}
\author{Li \surname{Cheng}$^{2}$}
\author{Yizhi \surname{Hu}$^{1}$}
\author{Junjie \surname{Gao}$^{1}$}
\author{Xiaolong \surname{Zou}$^{2}$}
\author{Xiaobin \surname{Chen}$^{1,3}$}
\email{Email:chenxiaobin@hit.edu.cn}

\affiliation{
$^1$School of Science, State Key Laboratory on Tunable Laser Technology and Ministry of Industry and Information Technology Key Lab of Micro-Nano Optoelectronic Information System, Harbin Institute of Technology, Shenzhen, Shenzhen 518055, China\\
$^2$Shenzhen Geim Graphene Center \& Shenzhen Key Laboratory of Advanced Layered Materials for Value-added Applications, Institute of Materials Research, Tsinghua Shenzhen International Graduate School, Tsinghua University, Shenzhen 518055, China\\
$^3$Collaborative Innovation Center of Extreme Optics, Shanxi University, Taiyuan 030006, China\\
}

\date{\today}
\begin{abstract}
Half metals, which are amenable to perfect spin filtering, can be utilized for high-magnetoresistive devices. However, available half metals are very limited. Here, we demonstrate that materials with intrinsic spin-valley-mismatched (SVM) states can be used to block charge transport, resembling half metals and leading to giant tunneling magnetoresistance. As an example, by using first-principles transport calculations, we show that ferromagnetic $1T$-VSe$_2$, 1\emph{T}-VS$_2$, and 2\emph{H}-VS$_2$ are such spin-valley-mismatched metals, and giant magnetoresistance of more than 99\% can be realized in spin-valve Van der Waals (VdW) junctions using these metals as electrodes. Owing to the intrinsic mismatch of spin states, the central-layer materials for the VdW junctions can be arbitrary nonmagnetic materials, in principle. Our research provides clear physical insights into the mechanism for high magnetoresistance and opens new avenues for the search and design of high-magnetoresistance devices.
\end{abstract}

\maketitle

The discovery of giant magnetoresistance ~\cite{Fert1988,Grunberg1989} has led to a surge of exploration on magnetoresistive devices, which have wide applications in magnetic sensing and data storage technologies~\cite{XuXiaodong_sci2018}. To enhance the sensitivity of magnetoresistive devices, high magnetoresistance (MR) is favored. Half metals are promising components for achieving large MR since only one spin channel is allowed to transport in half metals and thus perfect spin filtering can be naturally achieved in half-metal-based devices~\cite{Groot_PRL1983,sci1997,PRB1997}. However, progress in making use of half metals has been slow because of the difficulty in the manufacture of half metals and detailed control of the interfaces between the electrodes and the central-layer semiconductors or insulators~\cite{Karpan_PRL2007}.

The interface issues may be overcome by utilizing Van der Waals (VdW) materials. For example, if the insulator is replaced by 2D materials such as graphene, the MR devices can be less sensitive to interface details because of dramatically increased surface smoothness~\cite{Karpan_PRL2007}. With the presence of 2D magnetic materials CrI$_3$, spin-filter VdW junctions with giant tunneling MR (TMR) can be realized~\cite{XuXiaodong_sci2018}. However, half-metallic layered materials are rarely reported. Theoretically, only a few intrinsic layer-material half metals such as 2D ternary chalcogenides~\cite{ZhangShuqing_AFM2019} have been predicted. Thus far, searching for spin-valve VdW heterostructures with perfect low-conductance states remains a challenge.

In this work, we show that spin-valley-mismatched (SVM) materials, which are not half metals but have intrinsic mismatched spin-resolved valley states, can exhibit transport gaps in transport junctions. Being used as electrodes, such materials can block charge transport under the anti-parallel (AP) configuration, leading to giant TMR.

\begin{figure*}
\centering
    \includegraphics[width=0.8\textwidth]{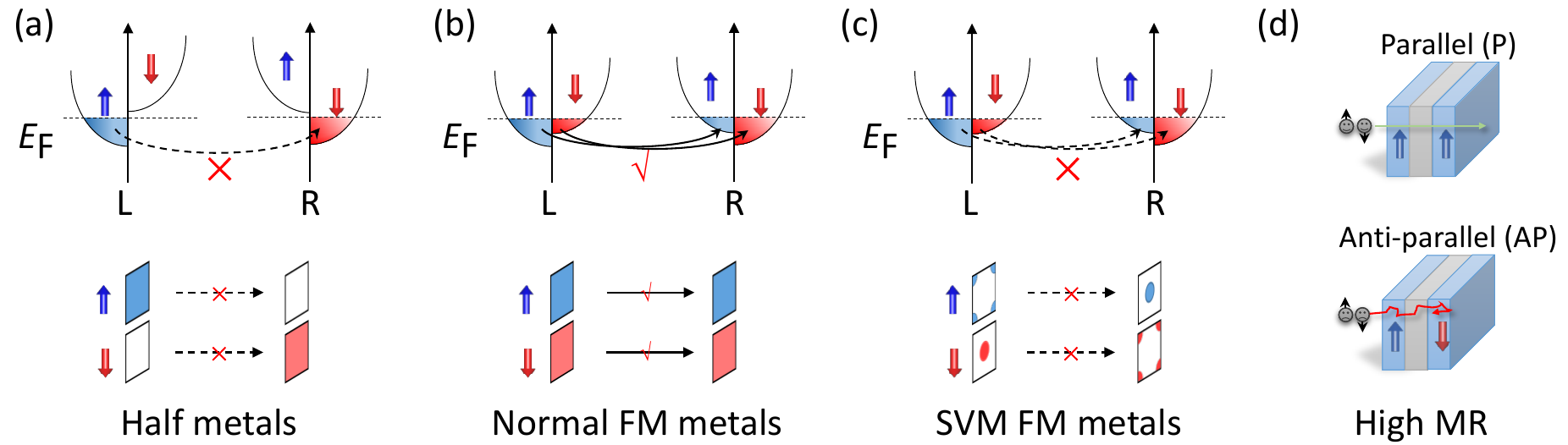}\\
  \caption{ Comparison of transmission under the AP configuration for using (a) half metals, (b) normal FM metals, and (c) SVM FM metals as the leads using spin-resolved density of states (upper panels) and Brillouin zones (lower panels). Shaded areas stand for occupied states. In (a), transport of electrons is prohibited due to spin mismatch of charge carriers.  In (b), the charge carriers are allowed to transport. By contrast, the transport between same-spin states is not allowed in (c) due to the mismatch of $\bf{k}$ vectors. (d) Based on the observations in (c), FM/Spacer/FM junctions using SVM FM metals may exhibit high MR.
  }\label{fig1}
\end{figure*}

To illustrate our idea, we first recall the mechanism of perfect spin filtering using half metals. It is well accepted that half metals can be used to obtain large MR~\cite{sci1997,PRB1997}. 
According to the Julliere model for spin tunneling systems, the optimistic/pessimistic MR is $\textrm{MR}_\textrm{o/p}=2\eta\eta'/(1\mp\eta\eta')$~\cite{JullierePLA1975,PRB1997}, where $\eta$ and $\eta'$ are the spin polarization of two leads. Thus, tunneling junctions with 100\%-spin-polarized leads are expected to exhibit divergent $\textrm{MR}_\textrm{o}$ or 100\% $\textrm{MR}_\textrm{p}$. As shown in Fig.~\ref{fig1}(a), there is only spin around the Fermi energy for a half metal. For a magnetic junction under the parallel (P) configuration, which means that the magnetization directions of the leads are parallel, electrons are allowed to transmit. For the AP configuration, the transporting spin channels in the left and right leads have different spins, and thus transportation is fully blocked. Giant MR of tunneling devices using half-metals as leads is demonstrated both experimentally and computationally~\cite{Wang_Npj2021,Taro_PRA2021,Kumar_APL2022,Lalrinkima_prb2023}.

Besides spin-filtering in half metals, we point out that high MR can be realized in SVM materials. As shown in Fig.~\ref{fig1}(b-c), same-spin tunneling is allowed when using normal ferromagnetic (FM) metals as leads but is blocked using SVM FM metals, for which the spin-up electrons are not allowed to transmit to the states with the same $\bf{k}$ vector because of spin mismatch nor to the states with the same spin because of $\bf{k}$-vector mismatch under the AP configuration~\cite{RycerzNP2007,cxbPRB2015,XuFuming2016,zhengXiaohong2Dmater2017,cxbPRB2019,NC2021_Tsymbal}. Therefore, giant MR can be realized in the SVM-materials-based spin-valve VdW junctions [Fig.~\ref{fig1}(d)]. To achieve SVM, the Kramers degeneracy should be lifted. Therefore, either the spatial inversion symmetry or the time-reversal symmetry should be broken. For magnetic materials, the time-reversal symmetry is naturally broken and the Kramers degeneracy is lifted. Therefore, one would expect to find SVM in various ferromagnetic materials.

For a demonstration, we choose the transition-metal dichalcogenides (TMDs) VSe$_2$ in the 1$T$ phase (denoted as $1T$-VSe$_2$), which has a hexagonal lattice and a layered structure [Fig.~2(a-b)], as the leads. TMDs are rich of valley-contrasting phenomena \cite{Vargiamidis_PRB2020}. From the band structure and Fermi surface of VSe$_2$ (Fig.~\ref{fig2}(c) and Sec. 1 of the Supplementary Material (SM)), we can see that it is a metal with separated spin-resolved states at the Fermi energy, resembling the Fermi surface of monolayer ferromagnetic VSe$_2$~\cite{Esters_prb2017,JPCC2019}.

The $1T$-VSe$_2$-based L/C/R transport model is shown in Fig.~\ref{fig2}(d). The ferromagnetic semi-infinite $1T$-VSe$_2$ serves as the leads, and the central scattering region contains six $1T$-VSe$_2$ layers as the buffer layers and a monolayer MX$_2$ (MoSe$_2$, WSe$_2$, MoS$_2$, and WS$_2$) in the 2$H$ phase (denoted as 2$H$-MoSe$_2$), where the X atoms are arranged in trigonal prismatic geometry \cite{ZouXL_AccChem2015}. The MX$_2$ serves as a tunneling barrier in the center. Fig.~\ref{fig2}(e) further shows the transporting channels of VSe$_2$ leads, for which the spin-up electrons are mainly concentrated around the K points and the spin-down electrons are mainly located around the M points. Except for the neighborhood of $\Gamma$ point, the spin-up and spin-down electrons of $1T$-VSe$_2$ do not overlap at the Fermi energy.
\begin{figure*}
\centering
  \centering
\includegraphics[width=0.8\textwidth]{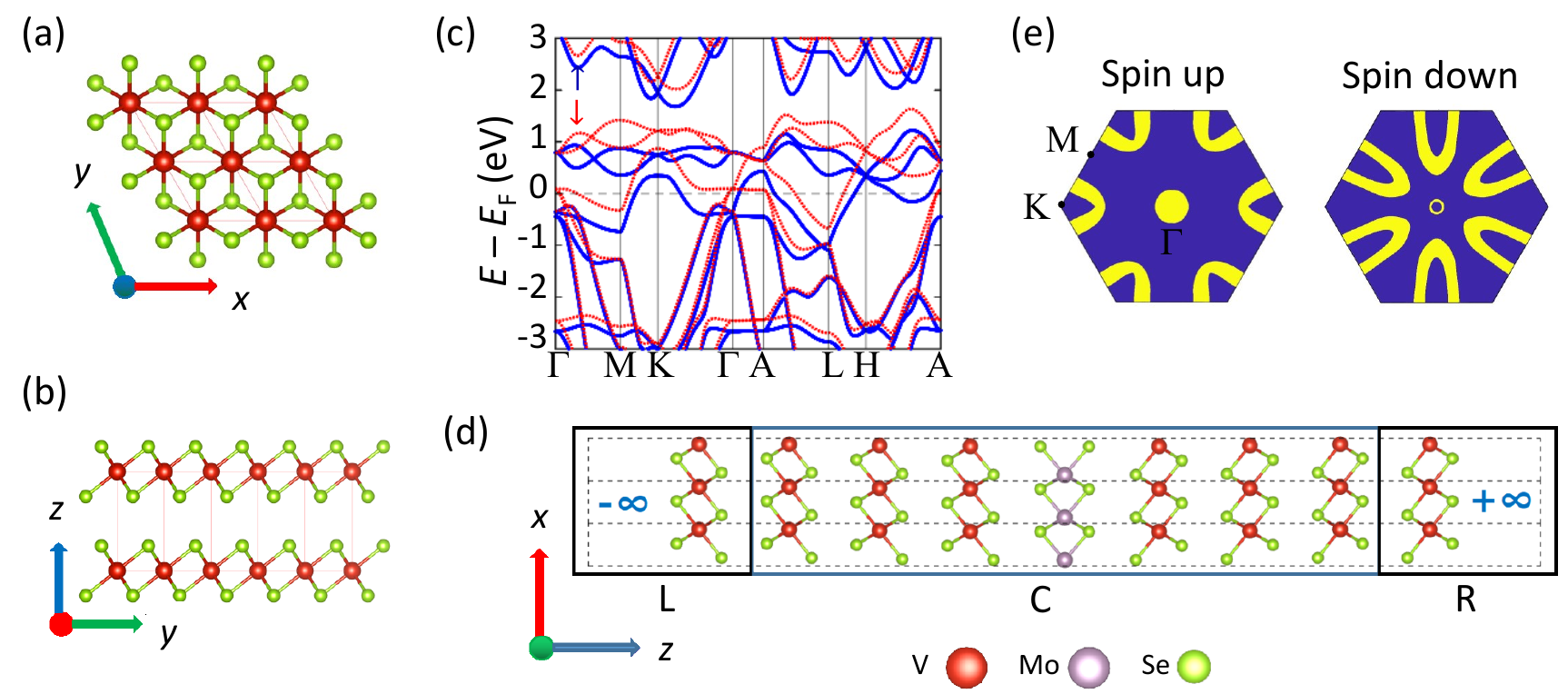}\\
\caption{(a) Top view and (b) side view of the crystalline structure of 1$T$-VSe$_2$. (c) Band structure and (d) the L/C/R model used for calculating transport properties of VSe$_2$-based magnetic tunneling junctions. Here, VSe$_2$/1-MoSe$_2$/VSe$_2$ is shown as an example. The system is infinite along the $z$ direction, and periodic along the $x$ and $y$ directions. (e) Spin-up (left panel) and spin-down (right panel) transporting channels of bulk VSe$_2$ illustrated in the first Brillouin Zone. Yellow color indicates the existence of transporting channels. }\label{fig2}
\end{figure*}

Methods for performing first-principles calculations can be found in Sec. 2 of the SM. In this work, we adopt both the optimistic and pessimistic versions of MR, which are defined as~\cite{Nanotechnology2022}
\begin{align}
{\rm{MR_o}} &= \frac{{{J^P} - {J^{AP}}}}{{{J^{AP}}}} \times 100\%,\\
{\rm{MR_p}} &= \frac{{{J^P} - {J^{AP}}}}{{{J^{P}}}} \times 100\%,
\end{align}
respectively, with $J^{P/AP}$ the total current density under $P/AP$ configuration.
For zero-bias cases, ${\rm{MR_{o(p)}}}=(T^P-T^{AP})/T^{AP(P)}\times 100\%$ can be obtained using the total transmission coefficients $T^{\alpha}$ under the $\alpha$ configuration ($\alpha=P,AP$). Note that the total current densities and the total transmission coefficients are both a summation over the spin-up and spin-down components as $J^{\alpha}=J_\uparrow^{\alpha}+J_\downarrow^{\alpha}$, $T^{\alpha}=T_\uparrow^{\alpha}+T_\downarrow^{\alpha}$.

In a spin-valve VdW junction, a central region is needed to better separate two FM regions with different magnetization directions and form a magnetic tunnel junction. For the case of monolayer MoSe$_2$ (1-MoSe$_2$), which has a large band gap of 1.5 eV~\cite{ACSnano2014}, the 1-MoSe$_2$ is stretched by 2\% to match the lattice of VSe$_2$.

\begin{figure}
\centering
  \centering
\includegraphics[width=0.45\textwidth]{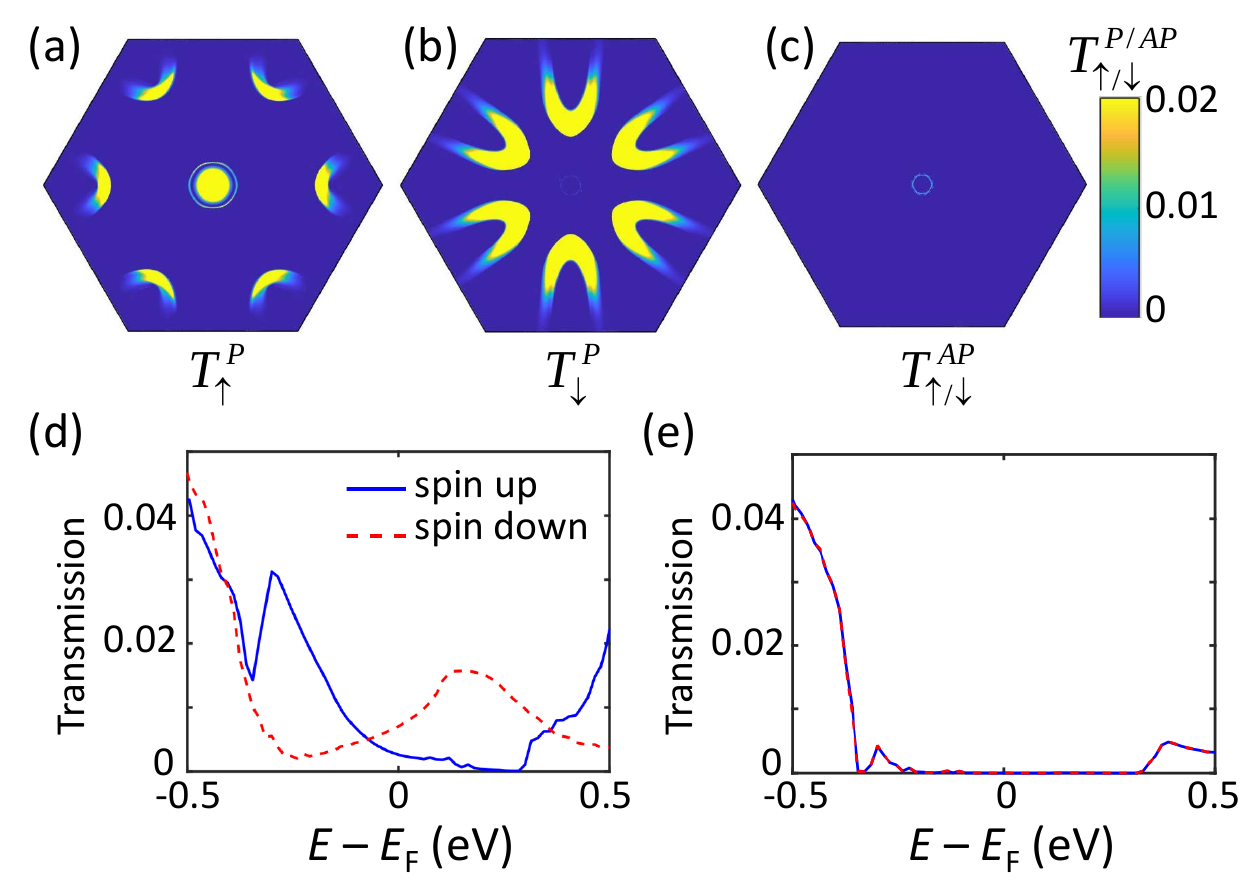}\\
  \caption{Zero-bias transport properties of the VSe$_2$/1-MoSe$_2$/VSe$_2$ junction. (a) spin-up and (b) spin-down transmission spectra under the P configuration and (c) spin-up/down transmission spectrum under the AP configuration. Transmission coefficients as a function of energy for spin-up (blue solid line) and spin-down (red dash line) electrons at zero bias under (d) P and (e) AP configurations.  }\label{fig3}
\end{figure}

Both spin- and $\bf{k}$-resolved transmission spectra of the VSe$_2$/1-MoSe$_2$/VSe$_2$ junction at zero bias are plotted in Fig.~\ref{fig3}(a-c). Overall, transmission coefficients are rather small, indicating a tunneling behavior. This feature is different from TMD junctions using metals as leads such as monolayer WSe$_2$ or MoS$_2$ sandwiched by Fe~\cite{Nanotechnology2022,Sanvito_PRB2014}, which are metallic due to the strong interaction between the metallic leads and the chalcogen atoms (S, Se). Specifically, for spin-up electrons under the P configuration, transmission distributes mainly around K corners and the $\Gamma$ point. While, for spin-down electrons, transmission is located in valleys around the M points. It is worth pointing out that distributions of the spin-up and spin-down transmission spectra are separated in the reciprocal space and their overlap is a small ring around the $\Gamma$ point. For the AP configuration, only a little transmission manifests around the $\Gamma$ point, consistent with the overlap of spin-up and spin-down transmission spectra under the P configuration [Fig.~\ref{fig3}(a-b)] and with the transporting channels shown in Fig.~\ref{fig2}(e). The observed transmission and filtering results are perfectly in line with our spin-valley-filtering theory.

Further, we plot the transmission coefficients of the VSe$_2$/1-MoSe$_2$/VSe$_2$ junction in Fig.~\ref{fig3}(d-e) to look into the spin-filtering characteristics. Under the P configuration, the transmission coefficients at the Fermi energy $E_F$ are about 0.0026 and 0.0071 for spin-up and spin-down electrons, respectively. Under the AP configuration, transmission coefficients of spin-up and spin-down electrons are almost the same and are vanishing around the Fermi energy, where a transport gap of 0.68 eV emerges. The zero-bias MR of VSe$_2$/1-MoSe$_2$/VSe$_2$ is $\rm{MR}_o~ (\rm{MR}_p) =1.85\times10^4\%~(99.5\%)$, which is much larger than magnetic tunneling junctions using metallic leads such as $\rm{MR}_o=3700\%$ in Fe/5-layer-MgO/Fe structures~\cite{Waldron2006}, $\rm{MR}_o=14.8\%$ in Ni/monolayer-WSe$_2$/Ni~\cite{Nanotechnology2022}, and spin-filter junctions such as Ni/Graphene/Ni~\cite{Karpan_PRL2007} and CrI$_3$-based junctions~\cite{XuXiaodong_sci2018}. Compared to other VSe$_2$-based junctions~\cite{YangWei_nanoscale2021}, our model is significantly simpler with clear physical insights. More zero-bias properties can be found in Sec. 3 of the SM.

\begin{table}[!hbp]
\caption{Calculated zero-bias MR of monolayer MX$_2$ (1-MX$_2$) or bilayer MX$_2$ (2-MX$_2$) sandwiched by VSe$_2$ or VS$_2$ leads.}
\begin{tabular}{c|c|cc}
  \hline
leads &  MX$_2$ &  $\rm{MR}_o$(\%) & $\rm{MR}_p$(\%) \\ \hline  \hline
      & 1-MoSe$_2$ & 1.85$\times10^4$ & 99.5 \\
       &1-WSe$_2$  & 2.21$\times10^3$ & 95.7 \\
          &1-MoS$_2$ & 1.09$\times10^4$ & 99.1 \\
  1\emph{T}-VSe$_2$ &  1-WS$_2$  & 2.49$\times10^3$& 96.1 \\
       & 2-MoSe$_2$ & 1.50$\times10^4$ & 99.3 \\
       &   2-WSe$_2$ & 2.08$\times10^3$ & 95.4 \\
       &  2-MoS$_2$& 3.94$\times10^3$ & 97.5 \\
       &  2-WS$_2$ & 2.07$\times10^3$ & 95.4 \\
        &  1-TiS$_2$ & 3.42$\times10^3$ & 97.2\\
        & Vaccum (5\AA) & 2.53$\times10^4$ & 99.6 \\
  \hline
  2\emph{H}-VS$_2$ & 1-MoS$_2$  & 1.26$\times10^8$ & 99.99 \\
    \hline
    1\emph{T}-VS$_2$ & 1-MoS$_2$  & 4.65$\times10^5$ & 99.98 \\
      \hline

\end{tabular}\label{table1}
\end{table}

\begin{figure}
\centering
  \centering
\includegraphics[width=0.49\textwidth]{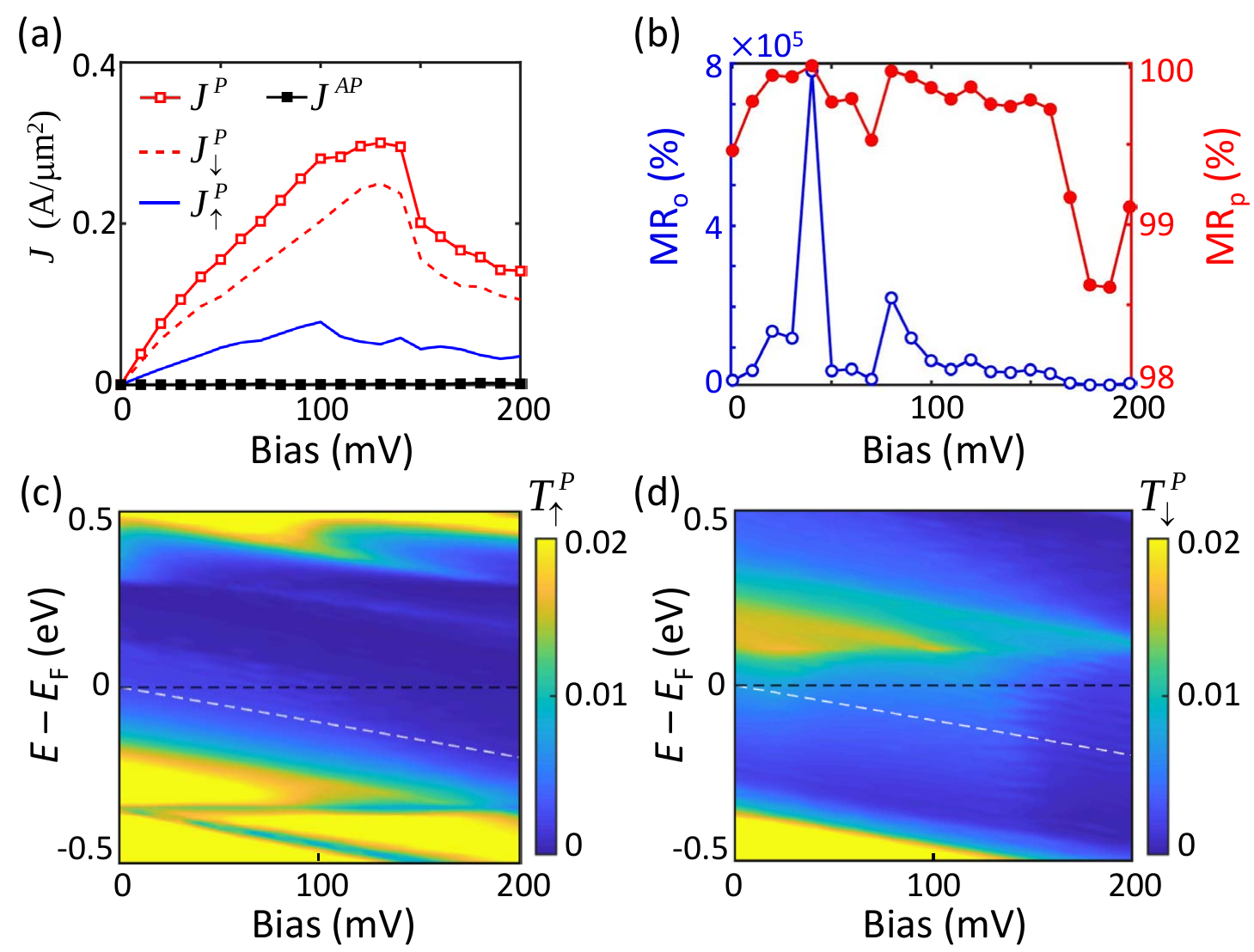}\\
  \caption{Transport properties of the VSe$_2$/1-MoSe$_2$/VSe$_2$ junction under a bias voltage. (a) Spin-up (blue solid line), spin-down (red dash line) current densities under the P configuration, and total current densities under the P (red solid line with empty circles) and AP (black line with filled circles) configurations. (b) MR$_o$ (left) and MR$_p$ (right) as a function of bias voltage. (c-d) Contour plot of the spin-up and spin-down transmission coefficients as functions of energy and bias under the P configuration. The black dash line and the white dash line indicate the bias window. }\label{fig:IV}
\end{figure}

Next, we explore the bias dependence of the MR. The calculated I-V curves and MR are depicted in Fig.~\ref{fig:IV} (a-b). Under P configuration, the spin-up and spin-down current densities show a linear dependence on the bias voltage up to about 100 and 130 mV, respectively. The major contribution comes from spin-down electrons at all biases, which agrees well again with Fig.~\ref{fig3}(d). As a consequence, the total current density under the P configuration also shows a peak value at a bias of about 130 mV and has the negative differential resistance (NDR), which has also reported in molecular tunnel junctions~\cite{HoW_PRL2008} and antiferroelectric multiferroic tunnel junctions~\cite{HanXiufeng_PRA2023}. Although oscillating with bias, $\rm{MR}_o$ ($\rm{MR}_p$) is larger than $7.0\times 10^3\%$ (98.6\%) within a bias voltage of 0 to 200 mV. The largest value of $\rm{MR}_o$ ($\rm{MR}_p$) is $7.8\times10^5\%$ (99.9\%), occurring at 40~mV. From the above results, VSe$_2$/1-MoSe$_2$/VSe$_2$ is a perfect carrier filter, showing robust and giant TMR under biases within 200 mV.

In addition, the NDR phenomenon can be ascribed to the competition between the decrease in transmission and the widening of bias window with the increase of bias, as indicated in Fig.~\ref{fig:IV}(c-d) and also Sec. 4 in SM~\cite{LiuNuo2016}. Although both the bias windows for spin-up and spin-down electrons increase with the bias, the transmission coefficients also vary with the bias. For spin-up electrons, the transmission coefficient has a significant decrease when the bias is larger than 100 mV. For spin-down electrons, a remarkable decrease in transmission coefficient occurs around 150 mV.

The energy barrier may change the MR~\cite{Vargiamidis_APL2014,Vasilopoulos_PRB2014}. However, according to our theory, various kinds of materials can act as the middle layer since the spin-valley-filtering effect is inherited from the leads. To validate this picture, we replace the central monolayer MoSe$_2$ with bilayer MoSe$_2$ and other TMD materials including the semiconducting monolayer and bilayer WSe$_2$, MoS$_2$, WS$_2$, 
and also the semi-metallic monolayer TiS$_2$.
 The calculated MR is shown in Tab.~\ref{table1}. Similarly to the case of monolayer MoSe$_2$, the other transport models exhibit giant $\rm{MR}_o$ ($\rm{MR}_p$) of $0.2-1.5\times10^4\%$ ($95\%-99\%$). In addition, we note that ferromagnetic VS$_2$ in the 1\emph{T} and 2\emph{H} phases, for which its monolayer form has been shown to have great potentials in lateral junctions~\cite{DaiY_pccp2018}, also have mismatched spin states~\cite{FuH_NJP2016}, and giant tunneling MR can be obtained in VS$_2$/1-MoS$_2$/VS$_2$ VdW junctions as shown in Tab.~\ref{table1}. (Details in the SM, Sec. 5-9.)

In our calculations, transverse periodic boundary conditions are set to reduce the computational workload \cite{PRR2021}. As a result, the central-layer materials are limited because the lattice mismatch between the electrodes and the central layers cannot be too large. However, note that the low-resistance state and transport gaps originate from the spin-valley mismatch of the electrodes. Therefore, the central-layer materials can, in principle, be arbitrary layered nonmagnetic materials. In addition, few-layer/monolayer VSe$_2$ and VS$_2$ are also SVM materials thanks to the 2D nature of VSe$_2$ and VS$_2$. With an appropriate choice of electrodes, high-MR devices can be fabricated based on few-layer or monolayer VSe$_2$ and VS$_2$, which demonstrate robust room-temperature ferromagnetism~\cite{NanoRes2022,AFM2023}. Based on our calculations, high MR still manifests under the charge density wave order phase (Sec. 10 in SM).

In summary, we proposed a new mechanism of spin-valley filtering that utilizes VdW ferromagnetic metals with intrinsic SVM states such as ferromagnetic 1\emph{T}-VSe$_2$, 1\emph{T}-VS$_2$, and 2\emph{H}-VS$_2$ for achieving giant MR in VdW heterojunctions. Using first-principles transport calculations, we predict that giant optimistic (pessimistic) MR larger than $1.8\times10^5$\% (99.5\%) can be realized in VSe$_2$/1-MoSe$_2$/VSe$_2$ under zero bias and the MR varies little under biases within 200~mV, with a peak value of $7.8\times10^5\%$ ($99.9\%$) is obtained at a bias of 40 mV. The VS$_2$-based VdW junctions also demonstrate MR larger than 99.9\%. These findings are heretofore unrecognized, and offer a new mechanism for realizing high MR in layered materials, guiding the design of new MR devices.

\begin{acknowledgments}
We would like to thank Prof. Wenhui Duan from Tsinghua University for helpful discussions. We gratefully acknowledge financial support by the National Natural Science Foundation of China (Grant No. 12074091) and the Shenzhen Science and Technology Program (Grant Nos. RCYX20221008092848063 and ZDSYS20230626091100001).

\end{acknowledgments}

%

\end{document}